# Diamond degradation in hadron fields[1]

*S. Lazanu[a], I. Lazanu[b], and E. Borchi[c]*

[a] Institute of Physics and Technology of Materials, POBox MG-7, Bucharest-Magurele, Romania

[b] University of Bucharest, POBox MG-11, Bucharest-Magurele, Romania

[c] Università di Firenze, Via S.Marta 3, 50139 - Florence, Italy

The energy dependence of the concentration of primary displacements induced by protons and pions in diamond has been calculated in the energy range 50 MeV - 50 GeV, in the frame of the Lindhard theory. The concentrations of primary displacements induced by protons and pions have completely different energy dependencies: the proton degradation is very important at low energies, and is higher than the pion one in the whole energy range investigated, with the exception of the $\Delta_{33}$ resonance region. Diamond has been found, theoretically, to be one order of magnitude more resistant to proton and pion irradiation in respect to silicon.

## 1. INTRODUCTION

The possible use of diamond detectors in high energy physics is due to its more appropriate material properties, in respect to, especially, silicon.

Diamond is, conceptually, a much simpler material to make detectors, as a result of its lack of p-n junction. A diamond detector works in an analogue way to an ionisation chamber: an electric field is applied between the 2 electrodes (sandwich configuration), and free carriers are generated at the passage of ionising radiation. The formed carriers induce signals to the electrodes.

Its resistivity, higher than $10^{14}$ $\Omega$cm, eliminated the need of any form of junction, conducing, this way, to the reduction of parallel shot noise in the readout electronics, in respect to silicon and GaAs.

Its low dielectric constant causes a small bulk and surface (interstrip) capacitance, which in turn reduces the series noise at the first stage of the signal amplifier. This is particularly important in high rate applications.

However, due to the high band gap of diamond, the energy to create an e-h pair is greater than for other semiconductor materials. So, for a 300 μm device, in the ideal situation, at the passage of a particle at minimum ionisation, the average signal is 13200 e-h pairs, (in comparison with 32400 in

---

silicon), while the most probable signal is about 10800 e-h pairs in diamond (and 21600 in Si). In reality, diamond detectors available today are far from ideal, and even a worse detected signal is obtained (about 40% of that generated).

Diamond is intrinsically very fast, particularly suitable for high rate operation. It is also an excellent thermal conductor, quality that offers advantage for large heat dissipation from the read-out electronics, and has excellent mechanical strength, that, coupled with low atomic number can reduce multiple Coulomb scattering.

It is its expected extreme radiation hardness that is the primary reason for considering diamond as a material for vertex detectors. Its outstanding properties (fast read-out with negligible noise, very good thermal management, reduction of multiple scattering) are shadowed by its (even in the ideal case) reduced signal.

The radiation resistance of diamond detectors has been experimentally confirmed for photons and electrons, and the tests for other particles are in progress.

In the present paper, the behaviour of diamond in hadron fields is presented, as the concentration of primary defects (CPD) induced by irradiation. Proton and pion degradations are calculated and compared, in the energy range 50 MeV - 50 GeV.

## 2. MECHANISMS OF DAMAGE AND RELEVANT QUANTITIES

In the projectile energies considered in this work, the hadron interacts dominantly with the nucleus, essentially unscreened, producing displacement defects in the diamond lattice. In the interaction process, one or more light energetic particles are produced. The light particles are supposed to escape from the diamond material, considered as a thin layer, without undergoing a second collision, while recoil nuclei loose all their energy in the lattice, by ionisation and atomic motion. They produce this way bulk damage, by subsequent collisions with other atoms.

As we shown before [1], the relevant quantity for the evaluation of radiation effects is the concentration of primary defects, produced by the unit particle fluence. This is calculated as the sum of the concentrations of defects resulting from all interaction processes, and all characteristic mechanisms corresponding to each interaction process.

The concentration of primary radiation induced defects (CPD) has been calculated as:

$$CPD = \frac{n(E)}{\Phi(E)}$$

with:

$$n(E) = \frac{N}{2E_d} \Phi(E) \sum_k \int d\Omega \sum_i \frac{d\sigma_i}{d\Omega} L(E_{Ri})$$

where: $E$ is the kinetic energy of the incident hadron; $N$ - atomic density of the target material; $E_d$ - threshold energy for displacements in the lattice; $\Phi(E)$ - the pion fluence in the primary beam; $E_{Ri}$ - recoil energy of the residual nucleus in the interaction $k$ ($k$= elastic, absorption and inelastic if the hadron is a pion, and $k$ = elastic, inelastic if it is a nucleon, respectively), in the interaction mechanism $i$, having a $d\sigma_i / d\Omega$ - differential cross section; $L(E_{Ri})$ - Lindhard factor describing the partition between ionisation and displacements. This factor describing the energy partition has been calculated in [2].

The contribution of each channel to the total concentration of defects depends on the probability of interaction and on the kinematics of the process, reflected in the recoil energy of the residual nucleus.

For monoatomic materials, the CPD is proportional to the non-ionisation energy loss (NIEL) [3], a physical quantity historically used for the characterisation of lattice degradation in particle fields.

## 3. HADRON INTERACTIONS

The interaction between a hadron and a nucleus can be either an elastic or an inelastic event.

In an elastic scattering process of interaction, symbolically represented by:

$h + Nucleus \rightarrow h + Nucleus$

the hadron does not excite the target nucleus.

The inelastic hadron - nucleus scattering includes all reactions of the type:

$h + Nucleus \rightarrow a_1 + a_2 + ... + a_n + Residual\ Nucleus$

where the reaction products $a_1, a_2, a_n$ can be proton, neutron, deuteron, other particles or light nuclei.

When the kinetic energy of the incident particle exceeds the threshold energy of 140 MeV, secondary pions could also be produced.

If the inelastic process is produced by nucleons, the identity of the incoming projectile is lost, and the creation of the secondary particles is associated with energy exchanges which are of the order of MeV or larger.

For pion - nucleus processes, a characteristic interaction is possible: the pion can disappear as a real particle within the nucleus, by absorption. In these calculations, the absorption process is considered separately. Absorption on a single nucleon is kinematically prohibited, and the simplest process is on two nucleons. Absorption on more nucleons is also possible.

The interaction of pions with nucleons and nuclei at kinetic energies comparable to the pion rest mass is dominated by the delta resonance production, with spin and isospin 3/2. At higher energies, other nucleons and resonances could be produced, but with much less importance.

Since the inelastic and absorption collisions are considerably more complex than elastic scattering, special reaction models must be developed for their analysis.

For the concrete calculations, the available experimental data have been used, and also different phenomenological approximations for their interpolations / extrapolation.

Rutherford and nuclear elastic scattering of hadrons on nuclei cannot be treated separately due to the strong interference mechanisms [4]. The data from reference [5] and [6-9] have been used respectively for pion and proton carbon differential cross sections, and they have been extended at other energies in the frame of a simple optical model, both for pions and for protons.

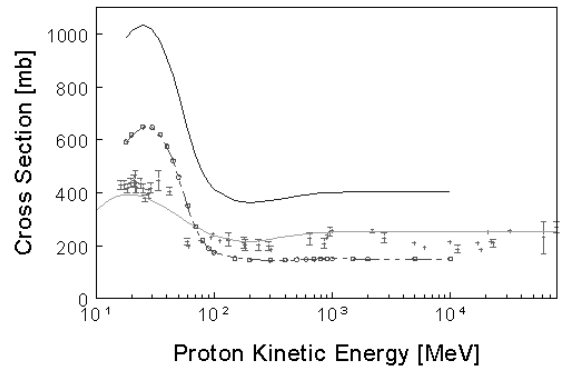

Figure 1. Energy dependence of proton - carbon cross sections: total, elastic and inelastic. The lines represent parametrisations of the data.

In Figures 1 and 2, the energy dependencies of the proton - carbon and pion - carbon cross sections are shown, respectively.

For proton - carbon interactions, the elastic cross sections are from [10], the inelastic ones (data) from [11], and the continuous curve represent the parametrisation from reference [12].

For positive pions - carbon interactions, the data are from [5] for the elastic cross sections, from [11, 13-14] for reaction, and from [13,15] for absorption respectively. The continuous lines are best fits of the data, and have been used to extrapolate / interpolate the values of the cross sections at energies of interest. The inelastic cross sections are obtained as differences between reaction and absorption ones, and are represented as dashed lines in Fig. 2.

The energy dependence of these cross sections, for proton and pion interactions with the carbon nucleus present very different behaviours: the proton - nucleus cross section decreases with the increase of the projectile energy, and has a minimum at low energies, while the pion - nucleus cross sections present for all processes a large maximum, at about 160 MeV, which reflects the resonance structure of the interaction.

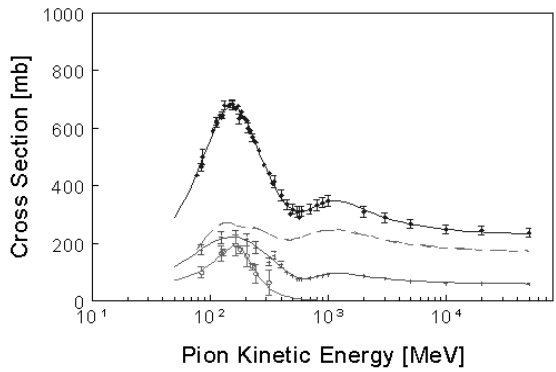

Figure 2. Energy dependence of pion - carbon cross sections: up to down the curves represent total, inelastic, elastic and absorption cross sections. The lines are the best fits of the data.

Both the pions and the nucleons interact strongly with the nucleus. It is important to compare nucleon - nucleus and pion - nucleus reactions. The dominant excitation modes directly affect the distribution of secondary fragments, which deposit ionisation energies in the host material. In reference [16], this study was been done. It can be seen that far from the delta resonance, the behaviour of pions and protons is somehow similar, and a scaling procedure could be applied from one particle to another, while in the $\Delta_{33}$ energy range this is not possible.

From the point of view of the isospin, the proton and the neutron are states of the nucleon, and $\pi^+$, $\pi^-$ and $\pi^0$ are different charge states of an unique system. The isospin symmetry is exact only for the strong interaction. In reality, this symmetry is broken by the electromagnetic interaction, which is manifested in the difference in mass between these particles, and in the different interaction probabilities of each particle with other systems.

For nucleon - nucleus interaction, above 100 MeV, the Coulomb interaction is not important, and the characteristics of proton - nucleus and neutron - nucleus reactions are very similar. Below 100 MeV, the cross sections for neutrons are somewhat larger that the corresponding proton cross sections, while the isotopic distributions of the residual nuclei from proton and neutron interactions are very similar. At around 20 MeV or below, the proton reaction cross section decreases rapidly because of the proton nucleus Coulomb barrier. This barrier does not apply to neutrons.

The interaction of $\pi^0$ with nuclei do not present interest for these studies. For $\pi^+$ and $\pi^-$, the differences in the interactions are due to the Coulomb barrier. Two aspects are relevant: the differences in the values of the cross sections, effect more important at low energies, and an energy shift of the maximum cross section in the region of the $\Delta_{33}$ resonance region. At high energies, far from the resonance, these effects decrease, and than become unimportant. The behaviour of the pion - nucleus interaction do not differ significantly for that of other hadrons.

## 4. MODEL CALCULATIONS, RESULTS AND DISCUSSIONS

Elastic and Rutherford contributions to the CPD have been treated together, as specified before.

The main difficulty is related to the inelastic interaction, due to the multitude of open channels, corresponding to possible final states. In order to obtain an estimation of the average recoil energies, some simplifying assumptions concerning this interactions have been made, starting from the available experimental data.. Both for pions and protons, the knock-out interaction has been considered separately, using the data from [13, 17-18] for pions, and from [10] for protons respectively. The rest of the channels have been considered as equivalent to the interaction on an effective number of nucleons.

Particle generation has been neglected.

In the case of pions, an important contribution is given by absorption, in the energy range of the delta resonance, while above 1 GeV we considered it as negligible. The simplest absorption mechanism involves two nucleons. We supposed that the absorption on quasideuteron is the dominant two nucleon mechanism. The experimental data show that in carbon the contribution of three-body absorption is also important, but the absorption on four or more nucleons is negligible [15]. The rest can be attributed to other mechanisms, especially to final state interactions, and has been treated statistically. Details could be found in [2].

Both for protons and pions, the energy dependence of the CPD follows the energy dependence of the cross sections.

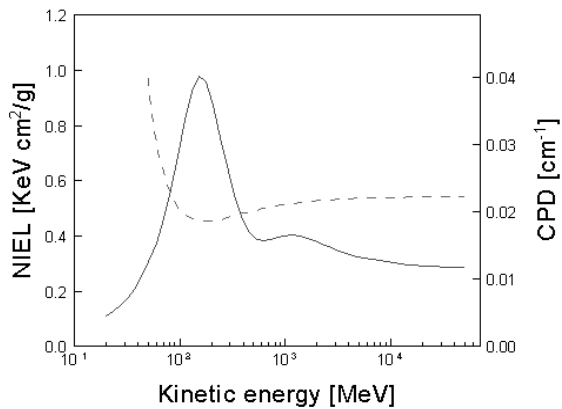

Figure 3. CPD - right scale, and NIEL - left scale produced by protons (dashed line), and pions (continuous line), in diamond.

The recoil energy partition between ionisation and displacements has been considered in the frame of the Lindhard theory.

The results obtained for the energy dependence of the CPD and NIEL produced by protons and pions are represented on the same graph in Figure 3. The CPD for protons present an abrupt decrease at low energies, followed by a minimum and at higher energies, by a plateau. For pions, there exists a large maximum in the region of the $\Delta_{33}$ resonance. The minimum for proton degradation and respectively the maximum for pion one are in the same energetic range.

Comparing the present results for the protons and pions degradation in diamond with the corresponding ones in silicon (for protons, see reference [3], for pions, reference [19]), the diamond proves to be one order of magnitude more resistant to radiation.

## 5. SUMMARY

The energy dependence of the CPD and NIEL produced by protons and pions in diamond have been calculated, in the energy range 50 MeV - 50 GeV.

Pions produce a higher degradation in diamond, in respect to protons, only in the $\Delta_{33}$ resonance energy range.

Diamond has been found one order of magnitude more resistant to proton and pion irradiation in respect to silicon.

Experimental measurements are necessary to validate the present calculations.